\documentclass{article}
\usepackage{mathrsfs}
\usepackage{amsfonts}
\usepackage{bbm}
\usepackage{amsmath, amsthm, amssymb}
\usepackage{color,calc}
\usepackage{hyperref}

\newcommand{\om}{\omega}

\begin{document}
 \vspace{15pt}

\begin{center}{\Large \bf The generalized Kupershmidt deformation for constructing new integrable systems from integrable bi-Hamiltonian systems}
\end{center}
\begin{center}
{\it Yuqin Yao\footnote{Corresponding author:
yqyao@math.tsinghua.edu.cn } and Yunbo
Zeng\footnote{yzeng@math.tsinghua.edu.cn} }
\end{center}
\begin{center}{\small \it$^{1)}$Department of
  Applied Mathematics, China Agricultural University, Beijing, 100083, PR China\\
 $^{2)}$Department of Mathematical Science,
Tsinghua University, Beijing, 100084 , PR China}
\end{center}

\vskip 12pt { \small\noindent\bf Abstract}
 {Based on the Kupershmidt deformation for any integrable bi-Hamiltonian systems presented in [4], we propose
 the generalized Kupershmidt deformation to construct new systems from integrable bi-Hamiltonian systems, which
 provides a nonholonomic perturbation of the bi-Hamiltonian systems. The generalized Kupershmidt deformation  is
  conjectured to preserve integrability. The conjecture is verified
  in a few representative cases: KdV equation, Boussinesq equation, Jaulent-Miodek equation and Camassa-Holm
  equation. For these specific cases, we present a general procedure to convert the
 generalized Kupershmidt deformation into
  the integrable Rosochatius deformation of soliton equation with self-consistent
  sources, then to transform it into a $t$-type bi-Hamiltonian system. By using this  generalized
   Kupershmidt deformation some new integrable systems are derived. In fact, this
generalized Kupershmidt deformation also
  provides a new method to construct the integrable Rosochatius deformation
  of soliton equation with self-consistent sources.}
\vskip 10pt

\section{Introduction}

It is known that one can construct a new integrable system starting
from a bi-Hamiltonian system. Fuchssteiner and Fokas showed
\cite{ff} that compatible symplectic structures lead in natural way
to hereditary symmetries, which provides a method to construct a
hierarchy of  exactly solvable evolution equations. Olver and
Rosenau \cite{OR} demonstrated that most integrable bi-Hamiltonian
systems are governed by a compatible trio of Hamiltonian structures,
and their recombination leads to integrable hierarchies of nonlinear
equations.

Recently for the following KdV6 equation or nonholonomic deformation
of KdV equation derived in \cite{kdv6}
\begin{subequations}
\label{eqns:i1}
 \begin{align}
&u_{t}=u_{xxx}+6uu_{x}-\omega_{x},\\
& \omega_{xxx}+4u\omega_{x}+2u_{x}\omega=0,
 \end{align}
\end{subequations}
Kupershmidt  found \cite{kd} that (\ref{eqns:i1}) can be converted
into
\begin{subequations}
\label{eqns:i2}
 \begin{align}
 &u_{t}=B_{1}(\frac{\delta H_{3}}{\delta u})
 -B_{1}(\omega),\label{eqns:i2a}\\
& B_{2}(\omega)=0,\label{eqns:i2b}
 \end{align}
\end{subequations}
where
\begin{equation}
\label{eqns:i3}
B_{1}=\partial=\partial_{x},~B_{2}=\partial^{3}+2(u\partial+\partial
u)
\end{equation}
are the two standard Hamiltonian operators of the KdV hierarchy
 and
$H_{3}=u^{3}-\frac{u_{x}^{2}}{2}.$ In general, for a bi-Hamiltonian
system
\begin{equation}
\label{eqns:i4} u_{t_{n}}=B_{1}(\frac{\delta H_{n+1}}{\delta
u})=B_{2}(\frac{\delta H_{n}}{\delta u}),
\end{equation}
the ansatz (\ref{eqns:i2}) provides a nonholonomic deformation of
bi-Hamiltonian systems\cite{kd}: $$u_{t_{n}}=B_{1}(\frac{\delta
H_{n+1}}{\delta u})
 -B_{1}(\omega),$$
\begin{equation} B_{2}(\omega)=0
 \end{equation}
which is called as Kupershmidt deformation of bi-Hamiltonian
systems. This deformation is conjectured to preserve integrability
and the conjecture is verified in a few representative cases in
\cite{kd}.

In \cite{yz0}, we showed that the Kupershmidt  deformation
(\ref{eqns:i2}) of KdV equation is equivalent to the integrable
Rosochatius deformation of KdV equation with self-consistent
sources, and constructed the bi-Hamiltonian structure for the
Kupershmidt
 deformation of KdV equation
(\ref{eqns:i2}). The conjecture is then proved in \cite{kkv} that
the Kupershmidt deformation of a bi-Hamiltonian system is itself
bi-Hamiltonian.

 Rosochatius found that it would still keep the
integrability to add a potential of the sum of inverse squares of
the coordinates to that of the Neumann system\cite{a1}. The deformed
system is called as Neumann-Rosochatius system. Then the Rosochatius
deformation of Garnier system, Jacobi system and many constrained
flows of soliton equations were constructed in \cite{a3,a4,a5}. This
Rosochatius-type integrable systems  have important physical
applications\cite{a6,a7,a8}. However, these Rosochatius deformations
are limited to few finite-dimensional integrable Hamiltonian
systems(FDIHS). Recently, in\cite{yz} we proposed a systematic
method for generalizing the integrable Rosochatius deformation for
FDIHS to integrable Rosochatius deformation for infinite-dimensional
integrable equations. Many integrable Rosochatius deformations of
soliton equations with self-consistent sources and their Lax
representations were presented in \cite{yz,yz1}.

In present paper, based on the Kupershmidt deformation (5), we
propose the generalized Kupershmidt deformation (GKD) to construct
new systems from integrable bi-Hamiltonian systems which provides a
nonholonomic perturbation of the bi-Hamiltonian systems. The
generalized Kupershmidt deformation  is
  conjectured to preserve integrability.  Although it is
difficult to prove the integrability in general, the conjecture can
be verified in many specific cases. Using KdV equation, Boussinesq
equation, Jaulent-Miodek equation and Camassa-Holm equation as
examples, we present a general procedure to show how to convert
these generalized Kupershmidt deformations into the Rosochatius
deformations of soliton equations with self-consistent sources.
These Rosochatius deformations of soliton equations with
self-consistent sources possess the Lax representations, which are
easy constructed by using the systematic method in \cite{yz,yz1},
and their stationary equations can be shown to be finite-dimensional
integrable Hamiltonian systems in the Liouville's sense
\cite{yz,yz1,yz2}. Furthermore, for the specific Rosochatius
deformations of soliton equations with self-consistent sources there
is a general procedure to transform it into a $t$-type
bi-Hamiltonian system by introducing the Jacobi-Ostrogradiski
coordinates and taking the spacial variable $x$ as the evolution
parameter according to\cite {h1,h2,h3,h4, yz0}. These facts imply
the integrability of the Rosochatius deformations of soliton
equations with self-consistent sources. Indeed  the generalized
Kupershmidt deformation also provides a new method for obtaining the
Rosochatius deformation of soliton equation with self-consistent
sources, which is quite different from the method presented in
\cite{yz,yz1}.

  In section 2, we propose the the generalized Kupershmidt
  deformation (GKD) of bi-Hamiltonian system, by using GKD of KdV hierarchy to illustrate the
  formulae. Section 3 is devoted to a new integrable system obtained from the GKD of Camassa-Holm
  equation and shows how to
   transform the GKD of Camassa-Holm equation into the integrable Rosochatius deformation of
Camassa-Holm equation with self-consistent sources. Section 4 treats
the GKD of Boussinesq equation. The last
  section presents the GKD of
  Jaulent-Miodek hierarchy, and demonstrate how to convert the
  GKD of a integrable system into a
  bi-Hamiltonian system with $t-$type Hamiltonian operator by taking $x$ as
  evolution parameter and $t$ as 'spatial' variable.

\section{The generalized Kupershmidt deformation of bi-Hamiltonian systems}
Consider a hierarchy of soliton equations with bi-Hamiltonian
structure
\begin{equation}
\label{eqns:0} u_{t_{n}}=B_{1}(\frac{\delta H_{n+1}}{\delta
u})=B_{2}(\frac{\delta H_{n}}{\delta u}),
\end{equation}
where $B_1$ and $B_2$ are two standard Hamiltonian operators. The
associated spectral problem reads
\begin{equation}
\label{eqns:00} L(u)\phi=\lambda\phi.
\end{equation}
For the eigenvalue $\lambda$, it is easy to find that the
variational derivative of $\lambda$
$$\frac{\delta \lambda}{\delta u}=f(\varphi).$$
 Assume that $\lambda_{j}, j=1,...N$ are $N$ distinct real eigenvalues of (7), we have
$$L\varphi_{j}=\lambda_{j}\varphi_{j},~j=1,2,\cdots,N,$$
and we denote
$$\frac{\delta \lambda_{j}}{\delta u}=\frac{\delta \lambda}{\delta u}|_{\lambda=\lambda_j}=f(\varphi_{j}).$$
Based on the  Kupershmidt deformation (5), we first generalize
Kupershmidt deformation  as follows
\begin{subequations}
\label{eqns:k2}
 \begin{align}
 &u_{t_{n}}=B_{1}(\frac{\delta H_{n+1}}{\delta u})
 -B_{1}(\sum_{j=1}^{N}\om_j),\label{eqns:k2a}\\
& (B_{2}-\alpha_{j}B_{1})(\om_j)=0,~j=1,2,\cdots, N,\label{eqns:k2b}
 \end{align}
\end{subequations}
where $\alpha_{j}$ are arbitrary constants, which also gives rise to
a nonholonomic deformation of bi-Hamiltonian systems (\ref{eqns:0})
similar to the integrable KdV6's type noholonomic deformation of
soliton equations (\ref{eqns:0}). So (\ref{eqns:k2}) provides a way
to construct new systems from the bi-Hamiltonian systems
(\ref{eqns:0}). Furthermore, observe that $\omega_{j}$ in (8a) are
at the same position as $\frac{\delta H_{n+1}}{\delta u}$, and the
eigenvalues $\lambda_j$ are also the conserved quantities for
(\ref{eqns:0}) as $H_n$, it is reasonable to take
$\om_j=\frac{\delta \lambda_{j}}{\delta u}$ and this setting is
compatible with (8b). So we finally propose the generalized
Kupershmidt deformation for a bi-Hamiltonian systems as follows
\begin{subequations}
\label{eqns:k1}
 \begin{align}
 &u_{t_{n}}=B_{1}(\frac{\delta H_{n+1}}{\delta u}
 -\sum_{j=1}^{N}\frac{\delta
\lambda_{j}}{\delta u}),\label{eqns:k1a}\\
&(B_{2}-\alpha_{j}B_{1})(\frac{\delta \lambda_{j}}{\delta
u})=0,~j=1,2,\cdots, N.\label{eqns:k1b}
 \end{align}
\end{subequations}
As the conjecture made in [4], it is reasonable to conjecture the
integrability of the new system (9). It seems that it is difficult
to prove the integrability in general. However it can be verified in
many specific cases. Using KdV equation, Boussinesq equation,
Jaulent-Miodek equation and Camassa-Holm equation as examples, we
will proceed to the general procedure to convert these generalized
Kupershmidt deformations into the Rosochatius deformations of
soliton equations with self-consistent sources. By using the
systematic method in \cite{yz,yz1}, it is easy to construct the Lax
representation for these Rosochatius deformations of soliton
equations with self-consistent sources, and to show their stationary
equations to be finite-dimensional integrable Hamiltonian systems in
the Liouville's sense \cite{yz,yz1}. Furthermore, following the
method proposed in \cite {h1,h2,h3,h4}, we have presented a general
procedure to transform the Rosochatius deformations of soliton
equations with self-consistent sources into a bi-Hamiltonian system
by introducing the Jacobi-Ostrogradiski coordinates and taking the
spacial variable $x$ as the evolution parameter in \cite {yz0}. We
will use the GKD of Jaulent-Miodek equation to illustrate the
general constructure. These facts imply the integrability of the
generalized Kupershmidt deformation of soliton equation. So the
deformation (\ref{eqns:k2}) and (\ref{eqns:k1}) provides a way to
construct new integrable systems from integrable bi-Hamiltonian
systems and to establish the integrable Rosochatius deformation of
soliton equation with self-consistent sources in different way from
that in \cite{yz,yz1}.

 We now use the KdV hierarchy to illustrate the procedure. Consider the Schr$\ddot{o}$dinger eigenvalue problem
\begin{equation}
\label{ad1} \phi_{xx}+(u-\lambda)\phi=0,
\end{equation}
the associated KdV hierarchy read
\begin{equation}
\label{ad2} u_{t_{n}}=B_{1}(\frac{\delta H_{n+1}}{\delta
u})=B_{2}(\frac{\delta H_{n}}{\delta u}), ~n=1,2,\cdots
\end{equation}
where
$$B_{1}=\partial=\partial_{x},~B_{2}=\partial^{3}+2(u\partial+\partial
u)$$$$H_{n+1}=\int b_{n+1}dx, ~~b_{n+1}=
-\frac{2}{2n+1}R^{n}u,~~R=-\frac{1}{4}\partial^{2}-u+\frac{1}{2}\partial^{-1}u_{x}.$$
It is easy to find that
$$\frac{\delta \lambda}{\delta u}=\varphi^{2}.$$
 For $N$ distinct  eigenvalues
$\lambda_{j}$, consider the  spectral problem
$$\varphi_{jxx}+(u-\lambda_{j})\varphi_{j}=0,~j=1,2,\cdots,N.$$
We have
$$\frac{\delta \lambda_{j}}{\delta u}=\varphi_{j}^{2}.$$

 For $n=2,~\alpha_j=\lambda_{j}$, (\ref{eqns:k2}) gives rise to the following new integrable generalized KdV6 equation
 \begin{subequations}
\label{eqns:i10}
 \begin{align}
&u_{t}=u_{xxx}+6uu_{x}-\sum_{j=1}^{N}\omega_{jx},\\&
\omega_{jxxx}+4u\omega_{jx}+2u_{x}\omega_j-\lambda_j\omega_{jx}=0,~j=1,2,\cdots,N.
 \end{align}
\end{subequations}
(\ref{eqns:k1b}) yields
$$2\varphi_{j}[\varphi_{jxx}+(u-\lambda_{j})\varphi_{j}]_{x}+6\varphi_{jx}[\varphi_{jxx}+(u-\lambda_{j})\varphi_{j}]=0,$$
which immediately gives rise to
$$\varphi_{jxx}+(u-\lambda_{j})\varphi_{j}=\frac{\mu_{j}}{\varphi_{j}^{3}},$$
where $\mu_{j},~j=1,2,\cdots,N$ are integral constants.
 When $n=2,~\alpha_{j}=\lambda_{j}$,(\ref{eqns:k1}) gives rise to the  following generalized Kupershmidt deformation of KdV equation
\begin{subequations}
\label{eqns:k20}
 \begin{align}
&u_{t}=\frac{1}{4}(u_{xxx}+6uu_{x})-\sum\limits_{j=1}^{N}(\varphi_{j}^{2})_{x},\\
&
\varphi_{jxx}+(u-\lambda_{j})\varphi_{j}=\frac{\mu_{j}}{\varphi_{j}^{3}},
~j=1,2,\cdots,N
 \end{align}
\end{subequations}
which is just the integrable Rosochatius deformation of KdV equation
with self-consistent sources (RD-KdVHSCS) presented in \cite{yz}.
When $\mu_{j}=0,~j=1,\cdots,N,$ (\ref{eqns:k20}) reduces to the KdV
equation with self-consistent sources\cite{ks}.  The Lax pair for
(\ref{eqns:k20}) was constructed by a systematic method in \cite{yz}
\begin{subequations}
\label{eqns:klax}
 \begin{align}
& \left(\begin{array}{c}
\psi_{1}\\
\psi_{2}\\
 \end{array}\right)_{x}=U\left(\begin{array}{c}
\psi_{1}\\
\psi_{2}\\
 \end{array}\right),~~U=\left(\begin{array}{cc}
0 & 1\\
\lambda-u & 0\\
 \end{array}\right),\label{eqns:klaxa} \\  \nonumber
& \left(\begin{array}{c}
\psi_{1}\\
\psi_{2}\\
 \end{array}\right)_{t}=V\left(\begin{array}{c}
\psi_{1}\\
\psi_{2}\\
 \end{array}\right),\\ \nonumber
 & V=\left(\begin{array}{cc}
-\frac{u_{x}}{4}& -\lambda+\frac{u}{2}\\
-\lambda^{2}-\frac{u}{2}\lambda-\frac{u_{xx}}{4}-\frac{u^{2}}{2}+\frac{1}{2}\sum\limits_{j=1}^{N}
\varphi_{j}^{2} & \frac{u_{x}}{4}\\
 \end{array}\right)\\
&
~~~~~-\frac{1}{2}\sum\limits_{j=1}^{N}\frac{1}{\lambda-\lambda_{j}}\left(\begin{array}{cc}
\varphi_{j}\varphi_{jx}& -\varphi_{j}^{2}\\
\varphi_{jx}^{2}+\frac{\mu_{j}}{\varphi_{j}^{2}}& -\varphi_{j}\varphi_{jx}\\
 \end{array}\right) \label{eqns:klaxb} .
 \end{align}
\end{subequations}
The stationary equation of (\ref{eqns:k20}) reduces to the
generalized integrable H$\acute{e}$non-Heiles system [14]. The
bi-Hamiltonian structure for (\ref{eqns:k20})was presented in [5]

\section{The new integrable system obtained from the Camassa-Holm equation}
The Camassa-Holm (CH) equation \cite{ff,ch1,ch3} read
\begin{equation}\label{c1}
m_{t}=B_1 \frac{\delta H_1}{\delta u}=B_2 \frac{\delta H_0}{\delta
u}=-2u_{x}m-um_{x},~~~m=u-u_{xx}+\omega
\end{equation}
where
$$
B_1=-\partial+\partial^{3},~B_2=m\partial+\partial m
$$
are the two standard Hamiltonian operators of the CH equation
 and
$$H_{0}=\frac{1}{2}\int(u^{2}+u_{x}^{2})dx,~H_{1}=\frac{1}{2}\int(u^{3}+uu_{x}^{2})dx.$$
The Lax pair for CH equation is
\begin{subequations}
\label{eqns:c2}
 \begin{align}
&\phi_{xx}=(\frac{1}{4}-\frac{1}{2}m\lambda)\phi,\\
& \phi_{t}=\frac{1}{2}u_{x}\phi-(\frac{1}{\lambda}+u)\phi_{x}.
 \end{align}
\end{subequations}
For $N$ distinct real eigenvalues $\lambda_{j}$, consider the
following spectral problem
\begin{equation}\label{c3}
\varphi_{jxx}=\frac{1}{4}\varphi_{j}-\frac{1}{2}m\lambda_{j}\varphi_{j},~j=1,2,\cdots,N.
\end{equation}
It is easy to find that
$$\frac{\delta \lambda}{\delta m}=\lambda\varphi^{2},~~~
\frac{\delta \lambda_{j}}{\delta m}=\lambda_{j}\varphi_{j}^{2}.$$

Then we obtain the following new nonholonomic deformation of the
Camassa-Holm equation from (8) with
$\alpha_{j}=\frac{1}{\lambda_{j}}$
\begin{subequations}
\label{eqns:c40}
 \begin{align}
 &m_{t}=
 -2u_{x}m-um_{x}+\sum_{j=1}^{N}[\omega_{jx}-\omega_{jxxx}],\label{eqns:c40a}\\
&
2m\omega_{jx}+m_{x}\omega_j+\lambda_j[\omega_{jx}-\omega_{jxxx}]=0,~j=1,2,\cdots,N.\label{eqns:c40b}
 \end{align}
\end{subequations}
 Take $\alpha_{j}=\frac{1}{\lambda_{j}}$
, Eq. (9) leads to the following generalized Kupershmidt deformation
of CH
\begin{subequations}
\label{eqns:c4}
 \begin{align}
 &m_{t}=B_1(\frac{\delta H_{1}}{\delta m}
 -\sum_{j=1}^{N}\frac{1}{\lambda_{j}}\frac{\delta \lambda_{j}}{\delta m})
 =-2u_{x}m-um_{x}+\sum_{j=1}^{N}[(\varphi_{j}^{2})_{x}-(\varphi_{j}^{2})_{xxx}],\label{eqns:c4a}\\
& (B_2-\frac{1}{\lambda_{j}}B_1)(\frac{1}{\lambda_{j}}\frac{\delta
\lambda_{j}}{\delta m})=0,~j=1,2,\cdots,N.\label{eqns:c4b}
 \end{align}
\end{subequations}
Then (\ref{eqns:c4b}) yields
$$2\varphi_{j}(\varphi_{jxx}+\frac{1}{2}\lambda_{j}m\varphi_{j}-\frac{1}{4}\varphi_{j})_{x}+
6\varphi_{jx}(\varphi_{jxx}+\frac{1}{2}\lambda_{j}m\varphi_{j}-\frac{1}{4}\varphi_{j})=0$$
which immediately gives rise to
$$\varphi_{jxx}=\frac{1}{4}\varphi_{j}-\frac{1}{2}m\lambda_{j}\varphi_{j}+\frac{\mu_{j}}
{\varphi_{j}^{3}}.$$  So
 Eq. (\ref{eqns:c4}) gives a new integrable system
\begin{subequations}
\label{eqns:c5}
 \begin{align}
 &m_{t}=-2u_{x}m-um_{x}+\sum_{j=1}^{N}[(\varphi_{j}^{2})_{x}-(\varphi_{j}^{2})_{xxx}],\label{eqns:c5a}\\
&
\varphi_{jxx}=\frac{1}{4}\varphi_{j}-\frac{1}{2}m\lambda_{j}\varphi_{j}+\frac{\mu_{j}}
{\varphi_{j}^{3}},~j=1,2,\cdots,N\label{eqns:c5b}
 \end{align}
\end{subequations}
which lax pair can be found by using the method in \cite{yz4}
\begin{subequations}
\label{eqns:c6}
 \begin{align}
& \left(\begin{array}{c}
\psi_{1}\\
\psi_{2}\\
 \end{array}\right)_{x}=U\left(\begin{array}{c}
\psi_{1}\\
\psi_{2}\\
 \end{array}\right),~~U=\left(\begin{array}{cc}
0 & 1\\
\frac{1}{4}-\frac{1}{2}\lambda m & 0\\
 \end{array}\right)\\ \nonumber
& \left(\begin{array}{c}
\psi_{1}\\
\psi_{2}\\
 \end{array}\right)_{t}=V\left(\begin{array}{c}
\psi_{1}\\
\psi_{2}\\
 \end{array}\right),\\ \nonumber
 & V=\left(\begin{array}{cc}
\frac{u_{x}}{2}& -\frac{1}{\lambda}-u\\
\frac{u}{4}-\frac{1}{4 \lambda}+\frac{mu\lambda}{2} &-\frac{u_{x}}{2}\\
 \end{array}\right)\\
&
~~~~~-\sum\limits_{j=1}^{N}\frac{\lambda\lambda_{j}}{\lambda-\lambda_{j}}\left(\begin{array}{cc}
\varphi_{j}\varphi_{jx}& -\varphi_{j}^{2}\\
\varphi_{jx}^{2}+\frac{\mu_{j}}{\varphi_{j}^{2}}& -\varphi_{j}\varphi_{jx}\\
 \end{array}\right)
 \end{align}
\end{subequations}
In fact  Eq. (\ref{eqns:c5}) can also be regarded as the RD-CHESCS.

\section{The generalized Kupershmidt deformation of Boussinesq equation}
For the following third-order eigenvalue problem\cite{b1}
\begin{equation}\label{b1}
L\phi=\phi_{xxx}+v\phi_{x}+(\frac{1}{2}v_{x}+w)\phi=\lambda\phi,
\end{equation}
the associated Boussinesq equation is
\begin{equation}\label{adb1}
\left(\begin{array}{c}
v\\
w\\
 \end{array}\right)_{t}=B_1\left(\begin{array}{c}
\frac{\delta H_{2}}{\delta
 v}\\
\frac{\delta H_{2}}{\delta
 w}\\
 \end{array}\right)=B_2\left(\begin{array}{c}
\frac{\delta H_{1}}{\delta
 v}\\
\frac{\delta H_{1}}{\delta
 w}\\
 \end{array}\right)=\left(\begin{array}{c}
2w_{x}\\
-\frac{2}{3}vv_{x}-\frac{1}{6}w_{xxx}\\
 \end{array}\right),
\end{equation}
where
$$
B_1=\left(\begin{array}{cc}
0 & \partial\\
\partial & 0\\
 \end{array}\right),$$$$B_2=\frac{1}{3}\left(\begin{array}{cc}
2\partial^{3}+2v\partial+v_{x} & 3w\partial+2w_{x}\\
3w\partial+w_{x} &
-\frac{1}{6}(\partial^{5}+5v\partial^{3}+\frac{15}{2}v_{x}
\partial^{2}+\frac{9}{2}v_{xx}\partial+4v^{2}\partial+v_{xxx}+4vv_{x})\\
 \end{array}\right)
$$
are the two standard Hamiltonian operators of the Boussinesq
equation
 and
$$H_{1}=\int wdx,~H_{2}=\int(\frac{1}{12}v_{x}^{2}-\frac{1}{9}v^{3}+w^{2})dx.$$
 For $N$
distinct real eigenvalues $\lambda_{j}$, consider the following
spectral problem and its adjoint spectral problem
\begin{subequations}\label{b1} \begin{align}
&\varphi_{jxxx}+v\varphi_{jx}+(\frac{1}{2}v_{x}+w)\varphi_{j}=\lambda\varphi_{j},\\
&\varphi_{jxxx}^{*}+v\varphi_{jx}^{*}+(\frac{1}{2}v_{x}-w)\varphi_{j}^{*}=-\lambda\varphi_{j}^{*},
~j=1,2,\cdots,N.
\end{align}
\end{subequations}
  We have
$$\frac{\delta \lambda_{j}}{\delta
v}=\frac{3}{2}(\varphi_{jx}\varphi_{j}^{*}-\varphi_{j}\varphi_{jx}^{*}),~
\frac{\delta \lambda_{j}}{\delta w}=3\varphi_{j}\varphi_{j}^{*}.$$
The generalized Kupershmidt deformed Boussinesq equation is given by
(9) with $\alpha_j=\lambda_j$ as follows
\begin{subequations}
\label{eqns:b2}
 \begin{align}
 &\left(\begin{array}{c}
v\\
w\\
 \end{array}\right)_{t}=B_1(\left(\begin{array}{c}
\frac{\delta H_{2}}{\delta
 v}\\
\frac{\delta H_{2}}{\delta
 w}\\
 \end{array}\right)
 -\sum_{j=1}^{N}\left(\begin{array}{c}
\frac{\delta \lambda_{j}}{\delta
v}\\
\frac{\delta \lambda_{j}}{\delta
w}\\
 \end{array}\right)),\label{eqns:b2a}\\
& (B_2-\lambda_{j}B_1)\left(\begin{array}{c} \frac{\delta
\lambda_{j}}{\delta
v}\\
\frac{\delta \lambda_{j}}{\delta
w}\\
 \end{array}\right)=0,~j=1,2,\cdots,N.\label{eqns:b2b}
 \end{align}
\end{subequations}

Set
$$f_{j}=\varphi_{jxxx}+v\varphi_{jx}+(\frac{1}{2}v_{x}+w)\varphi_{j}-\lambda_j\varphi_{j},$$
$$(B_2-\lambda_{j}B_1)\left(\begin{array}{c}
\frac{\delta \lambda_{j}}{\delta
v}\\
\frac{\delta \lambda_{j}}{\delta
w}\\
 \end{array}\right)=\left(\begin{array}{c}\mathscr{A}_j\\
\mathscr{B}_j\\
 \end{array}\right),~j=1,2,\cdots,N.$$
Direct calculation gives
\begin{subequations}\label{b3} \begin{align}
\mathscr{A}_j=&\varphi_{j}^{*}f_{jx}+2\varphi_{jx}^{*}f_{j}-(2\varphi_{jx}^{*}f_{j}+\varphi_{j}^{*}f_{jx})^{*},
\label{b3a}\\ \nonumber
which ~leads ~to~~~~~~~~~~~~~~~~~~~~~~~~~& \\
&f_{j}=\frac{\mu_{j}}{\varphi_{j}^{*2}},~~f_{j}^{*}=-\frac{\mu_{j}}{\varphi_{j}^{2}}.
\label{b3b}
\end{align}
\end{subequations}
Using (\ref{b3b}), we get
$$
\mathscr{B}_j=-\frac{1}{3}(2v\varphi_{j}+5\varphi_{jxx})(f_{j}^{*}+\frac{\mu_{j}}{\varphi_{j}^{2}})-
\frac{1}{3}(2v\varphi_{j}^{*}+5\varphi_{jxx}^{*})(f_{j}-\frac{\mu_{j}}{\varphi_{j}^{*2}})
-\frac{5}{6}
\varphi_{jx}^{*}(f_{j}-\frac{\mu_{j}}{\varphi_{j}^{*2}})_{x}$$$$-\frac{5}{6}
\varphi_{jx}(f_{j}^{*}+\frac{\mu_{j}}{\varphi_{j}^{2}})_{x}-\frac{1}{6}
\varphi_{j}^{*}(f_{j}-\frac{\mu_{j}}{\varphi_{j}^{*2}})_{xx}
-\frac{1}{6}
\varphi_{j}(f_{j}^{*}+\frac{\mu_{j}}{\varphi_{j}^{2}})_{xx}+\frac{2\mu_{j}}{3\varphi_{j}^{3}\varphi_{j}^{*3}}$$$$
[v\varphi_{j}^{2}\varphi_{j}^{*2}(\varphi_{j}^{*}-\varphi_{j})-\varphi_{j}^{*3}\varphi_{jx}^{2}+\varphi_{j}^{3}(\varphi_{jx}^{*2}-2\varphi_{j}^{*}\varphi_{jxx}^{*})+2\varphi_{j}^{*3}
\varphi_{j}\varphi_{jxx}]$$
\begin{equation}
\label{b4}=\frac{2\mu_{j}}{3\varphi_{j}^{3}\varphi_{j}^{*3}}
[v\varphi_{j}^{2}\varphi_{j}^{*2}(\varphi_{j}^{*}-\varphi_{j})-\varphi_{j}^{*3}\varphi_{jx}^{2}
+\varphi_{j}^{3}(\varphi_{jx}^{*2}
-2\varphi_{j}^{*}\varphi_{jxx}^{*})+2\varphi_{j}^{*3}
\varphi_{j}\varphi_{jxx}],
\end{equation}
which yields to $\mu_{j}=0.$ Thus,
   the generalized Kupershmidt
deformed Boussinesq equation (\ref{eqns:b2}) gives the following
integrable system
\begin{subequations}\label{b4} \begin{align}
&v_{t}=2w_{x}-3\sum_{j=1}^{N}(\varphi_{j}\varphi_{j}^{*})_{x},\\
&w_{t}=-\frac{1}{6}(4vv_{x}+v_{xxx})-\frac{3}{2}(\varphi_{jxx}\varphi_{j}^{*}-\varphi_{j}\varphi_{jxx}^{*}),\\
&\varphi_{jxxx}+v\varphi_{jx}+(\frac{1}{2}v_{x}+w)\varphi_{j}=\lambda_j\varphi_{j},\\
&\varphi_{jxxx}^{*}+v\varphi_{jx}^{*}+(\frac{1}{2}v_{x}-w)\varphi_{j}^{*}=-\lambda_j\varphi_{j}^{*},
~j=1,2,\cdots,N
\end{align}
\end{subequations}
which just is the Boussinesq equation with self-consistent sources
and has the following Lax representation \cite{b2}
\begin{subequations}\label{b5} \begin{align}
&L_{t}=[\partial^{2}+\frac{2}{3}v+\sum_{j=1}^{N}\varphi_{j}\partial^{-1}\varphi_{j}^{*},L]\\
&L\psi=(\partial^{3}+v\partial+\frac{1}{2}v_{x}+w)\psi=\lambda\psi,\\
&\psi_{t}=(\partial^{2}+\frac{2}{3}v+\sum_{j=1}^{N}\varphi_{j}\partial^{-1}\varphi_{j}^{*})\psi.
\end{align}
\end{subequations}

\section{The generalized Kupershmidt deformation of Jaulent-Miodek equation and its bi-Hamiltonian structure}
\subsection{The generalized Kupershmidt deformation of Jaulent-Miodek equation}
The JM eigenvalue problem reads\cite{j1}
\begin{equation}
\label{j1} \left(\begin{array}{c}
\psi_{1}\\
\psi_{2}\\
 \end{array}\right)_{x}=U\left(\begin{array}{c}
\psi_{1}\\
\psi_{2}\\
 \end{array}\right),~~U=\left(\begin{array}{cc}
0 & 1\\
-\lambda^{2}+\lambda q+r & 0\\
 \end{array}\right),
\end{equation}
the associated JM hierarchy is
$$\left(\begin{array}{c}
q\\
r\\
 \end{array}\right)_{t_{n}}=B_1\left(\begin{array}{c}
b_{n+2}\\
b_{n+1}\\
 \end{array}\right)=B_1\left(\begin{array}{c}
\frac{\delta H_{n+1}}{\delta
q}\\
\frac{\delta H_{n+1}}{\delta
 r}\\
 \end{array}\right)=B_2\left(\begin{array}{c}
\frac{\delta H_{n}}{\delta
q}\\
\frac{\delta H_{n}}{\delta
 r}\\
 \end{array}\right)$$
where $$B_1=\left(\begin{array}{cc}
0 & 2\partial\\
2\partial & -q_{x}-2q\partial\\
 \end{array}\right),~B_2=\left(\begin{array}{cc} 2\partial & 0\\
0 & r_{x}+2r\partial-\frac{1}{2}\partial^{3}\\
 \end{array}\right),$$$$\left(\begin{array}{c}
b_{n+2}\\
b_{n+1}\\
 \end{array}\right)=L\left(\begin{array}{c}
b_{n+1}\\
b_{n}\\
 \end{array}\right),~n=1,2,\cdots$$
 $$b_0=b_1=0,~b_2=-1,~H_{n}=\frac{1}{n-1}(2b_{n+2}-qb_{n+1}).$$
 For $N$ distinct real eigenvalues $\lambda_{j}$, from the spectral problem
$$\varphi_{1jx}=\varphi_{2j},~\varphi_{2jx}=(-\lambda_{j}^{2}+\lambda_{j} q+r)\varphi_{1j}$$
we have
$$\frac{\delta \lambda_{j}}{\delta q}=\frac{1}{2}\lambda_{j}\varphi_{1j}^{2},~
\frac{\delta \lambda_{j}}{\delta r}=\frac{1}{2}\varphi_{1j}^{2}.$$
 Similarly, the generalized Kupershmidt
deformation (9) with $\alpha_j=\lambda_j$ for JM hierarchy gives
rise to
\begin{subequations}
\label{eqns:j2}
 \begin{align}
 &\left(\begin{array}{c}
q\\
r\\
 \end{array}\right)_{t_{n}}=B_1(\left(\begin{array}{c}
\frac{\delta H_{n+1}}{\delta
q}\\
\frac{\delta H_{n+1}}{\delta
 r}\\
 \end{array}\right)
 +\sum_{j=1}^{N}\left(\begin{array}{c}
\frac{\delta \lambda_{j}}{\delta
q}\\
\frac{\delta \lambda_{j}}{\delta
r}\\
 \end{array}\right))
 ,\label{eqns:j2a}\\
& (B_2-\lambda_{j}B_1)\left(\begin{array}{c} \frac{\delta
\lambda_{j}}{\delta
q}\\
\frac{\delta \lambda_{j}}{\delta
r}\\
 \end{array}\right)=0,~j=1,2,\cdots,N.\label{eqns:j2b}
 \end{align}
\end{subequations}
The first equation in (31b) is an identity and second one in
(\ref{eqns:j2b}) yields
$$\varphi_{1j}(\varphi_{1jxx}-r\varphi_{1j}-\lambda_{j}q\varphi_{1j}+\lambda_{j}^{2}\varphi_{1j})_x
+3\varphi_{1jx}(\varphi_{1jxx}-r\varphi_{1j}-\lambda_{j}q\varphi_{1j}+\lambda_{j}^{2}\varphi_{1j})=0$$
which, by setting $\varphi_{2j}=\varphi_{1jx}$, leads to
$$\varphi_{2jx}=(-\lambda_{j}^{2}+\lambda_{j}
q+r)\varphi_{1j}+\frac{\mu_{j}}{\varphi_{1j}^{3}},
~j=1,2,\cdots,N.$$ Then the generalized Kupershmidt deformation with
 of JM equation (\ref{eqns:j2}) with $n=3$ gives
rise to the following integrable system
\begin{subequations}\label{j3} \begin{align}
&q_{t}=-r_{x}-\frac{3}{2}qq_{x}+2\sum_{j=1}^{N}\varphi_{1j}\varphi_{2j},\\
&r_{t}=\frac{1}{4}q_{xxx}-q_{x}r-\frac{1}{2}qr_{x}+\sum_{j=1}^{N}[2(\lambda_{j}-q)\varphi_{1j}\varphi_{2j}
-\frac{1}{2}q_{x}\varphi_{1j}^{2}],\\
&\varphi_{1jx}=\varphi_{2j},~\varphi_{2jx}=(-\lambda_{j}^{2}+\lambda_{j}
q+r)\varphi_{1j}+\frac{\mu_{j}}{\varphi_{1j}^{3}} ~j=1,2,\cdots,N
\end{align}
\end{subequations}
which just is  the integrable RD-JMESCS and has the Lax
representation (\ref{eqns:klaxa})  with  \cite{yz1}
$$U=\left(\begin{array}{cc}
0 & 1\\
-\lambda^{2}+\lambda q+r & 0\\
 \end{array}\right),$$
$$V=\left(\begin{array}{cc}
 \frac{1}{4}q_{x}& -\lambda-\frac{1}{2}q\\
\lambda^{3}-\frac{1}{2}q\lambda^{2}-(\frac{1}{2}q^{2}+r)\lambda+\frac{1}{4}q_{xx}-\frac{1}{2}qr &-\frac{1}{4}q_{x}\\
 \end{array}\right)
 $$
 $$+\frac{1}{2}\left(\begin{array}{cc}
 0& 0\\
\lambda\langle \Phi_{1},\Phi_{1}\rangle-\langle
\Lambda\Phi_{1},\Phi_{1}\rangle-q
\langle \Phi_{1},\Phi_{1}\rangle &0\\
 \end{array}\right)+\frac{1}{2}\sum\limits_{j=1}^{N}\frac{1}{\lambda-\lambda_{j}}
 \left(\begin{array}{cc}
 \phi_{1j} \phi_{2j}& - \phi_{1j}^{2}\\
 \phi_{2j}^{2}+\frac{\mu_{j}}{\phi_{1j}^{2}}&-\phi_{1j} \phi_{2j}\\
 \end{array}\right)$$

\subsection{The bi-Hamiltonian structure of  GKDJME}
In the following, we will show how to construct the bi-Hamiltonian
structure for the generalized Kupershmidt deformation of a
bi-Hamiltonian system. We follow the method in \cite{h1}-\cite{h4}
to construct the bi-Hamiltonian formalism with $t-$ type Hamiltonian
operator for GKDJME by taking $x$ as the evolution parameter and $t$
as the 'spatial' variable. We denote the inner product in
$\mathbb{R}^{N}$ by $\langle.,.\rangle$ and
$$\Phi_{i}=(\varphi_{i1},\varphi_{i2},\cdots,\varphi_{iN})^{T},~~i=1,2,~~\mu=(\mu_{1},\cdots,\mu_{N})^{T},~~
\Lambda=diag(\lambda_{1},\cdots,\lambda_{N}).$$ Eq.(\ref{j3}) can be
written as
\begin{subequations}\label{j4} \begin{align}
\left(\begin{array}{c}
q\\
r\\
 \end{array}\right)_{t}=B_1\left(\begin{array}{c}
\frac{1}{8}q_{xx}-\frac{3}{4}qr-\frac{5}{16}q^{3}+\frac{1}{2}\langle \Lambda\Phi_{1},\Phi_{1} \rangle\\
-\frac{1}{2}r-\frac{3}{8}q^{2}+\frac{1}{2}\langle \Phi_{1},\Phi_{1} \rangle\\
 \end{array}\right) \label{j4a}\\
\varphi_{1jx}=\varphi_{2j},~\varphi_{2jx}=-\lambda_{j}^{2}\varphi_{1j}+q\lambda_{j}\varphi_{1j}
+r\varphi_{1j}+\frac{\mu_{j}}{\varphi_{1j}^{3}}.\label{j4b}
\end{align}
\end{subequations}
Notices that Kernel of $B_1$ is
$(c_{1}+\frac{1}{2}qc_{2},c_{2})^{T}$, we may rewrite (\ref{j4a}) as
\begin{subequations}\label{j5}\begin{align}
&\frac{1}{8}q_{xx}-\frac{3}{4}qr-\frac{5}{16}q^{3}+\frac{1}{2}\langle
\Lambda\Phi_{1},\Phi_{1} \rangle=
c_{1}+\frac{1}{2}qc_{2},~-\frac{1}{2}r-\frac{3}{8}q^{2}+\frac{1}{2}\langle
\Phi_{1},\Phi_{1} \rangle=c_{2}\label{j5a}\\
&c_{1x}=\frac{1}{2}\partial_{t}(r+\frac{1}{4}q^{2}),~c_{2x}=\frac{1}{2}\partial_{t}q
.\label{j5b}\end{align}\end{subequations} By introducing
\begin{equation}
\label{j6} q_{1}=q,~~p_{1}=-\frac{1}{8}q_{x}\end{equation}  Eqs.
(\ref{j4b}) and (\ref{j5b}) give rise to the t-type Hamiltonian form
\begin{subequations}
\label{j7}
\begin{align}
&R_{x}=G_{1}\frac{\delta F_{1}}{\delta R},\label{j7a}\\ \nonumber  where~~~~~&\\
\nonumber &
R=(\Phi_{1}^{T},q_{1},\Phi_{2}^{T},p_{1},c_{1},c_{2})^{T},\\
\nonumber
&F_{1}=-4p_{1}^{2}-\frac{1}{16}q_{1}^{4}-\frac{1}{2}q_{1}^{2}c_{2}+q_{1}c_{1}-c_{2}^{2}
+\frac{3}{8}q_{1}^{2}\langle
\Phi_{1},\Phi_{1}\rangle-\frac{1}{2}q_{1}\langle
\Lambda\Phi_{1},\Phi_{1}\rangle \label{j7b}\\
&+\frac{1}{2}\langle \Phi_{2},\Phi_{2}\rangle+\frac{1}{2}\langle
\Lambda^{2}\Phi_{1},\Phi_{1}\rangle+c_{2}\langle
\Phi_{1},\Phi_{1}\rangle-\frac{1}{4}\sum_{j=1}^{N}\varphi_{1j}^{4}+\frac{1}{2}
\sum_{j=1}^{N}\frac{\mu_{j}}{\varphi_{1j}^{2}},\\
\nonumber and ~ the ~~&~t-type~ Hamiltonian ~operator~ G_{1}~ is
~given~ by\\ & G_{1}=\left(\begin{array}{cccc} 0 &
I_{(N+1)\times(N+1)}
&0 &0\\
-I_{(N+1)\times(N+1)}& 0 &0& 0\\ 0&0&0&\frac{1}{2}\partial_{t}\\
0&0&\frac{1}{2}\partial_{t}&0
 \end{array}\right)\label{j7c}.
\end{align}
\end{subequations}

The modified Jaulent-Miodek (MJM) eigenvalue problem reads
\cite{Fordy}
\begin{equation}
\label{mj1} \left(\begin{array}{c}
\tilde{\psi}_{1}\\
\tilde{\psi}_{2}\\
 \end{array}\right)_{x}=\tilde{U}(\tilde{u},\lambda)\left(\begin{array}{c}
\tilde{\psi}_{1}\\
\tilde{\psi}_{2}\\
 \end{array}\right),~~\tilde{U}=\left(\begin{array}{cc}
-\tilde{r} & \lambda\\
-\lambda+\tilde{q}  &\tilde{r} \\
 \end{array}\right),~\tilde{u}=\left(\begin{array}{c}
\tilde{r}\\
\tilde{q}\\
 \end{array}\right)
\end{equation}
the associated MJM equation is of the form
$$\tilde{u}_{t}=\left(\begin{array}{c}
\tilde{r}\\
\tilde{q}\\
 \end{array}\right)_{t}=\left(\begin{array}{c}
-\frac{1}{4}\tilde{q}_{xx}-\frac{1}{2}(\tilde{q}\tilde{r})_{x}\\
-2\tilde{r}\tilde{r}_{x}-\frac{3}{2}\tilde{q}\tilde{q}_{x}+\tilde{r}_{xx}\\
 \end{array}\right)=\tilde{B}_1\frac{\delta \tilde{H}_{2}}{\delta
\tilde{u}}$$
where $\tilde{B}_1=\left(\begin{array}{cc} \frac{1}{2}\partial & 0\\
0 & 2\partial\\
 \end{array}\right),~\tilde{H}_{2}=-\frac{1}{2}\tilde{q}_{x}\tilde{r}-\frac{1}{2}\tilde{q}\tilde{r}^{2}
 -\frac{1}{8}\tilde{q}^{3}.$

 We have
 $$\frac{\delta \lambda}{\delta \tilde{u}}=\left(\begin{array}{c}
\tilde{\varphi}_{1}\tilde{\varphi}_{2}\\
\frac{1}{2}\tilde{\varphi}_{1}^{2}\\
 \end{array}\right)$$
Then the Rosochatius deformation of MJM equation with
self-consistent sources (RD-MJMSCS) is defined as
\begin{subequations}
\label{mj1}
\begin{align}
&\left(\begin{array}{c}
\tilde{r}\\
\tilde{q}\\
 \end{array}\right)_{t}=\tilde{B}_{1}(\frac{\delta \tilde{H}_{2}}{\delta \tilde{u}}
 +\frac{\delta \lambda}{\delta \tilde{u}})=\tilde{B}_{1}\left(\begin{array}{c}
-\frac{1}{2}\tilde{q}_{x}-\tilde{q}\tilde{r}+\langle\tilde{\Phi}_{1},\tilde{\Phi}_{2}\rangle\\
-\frac{1}{2}\tilde{r}^{2}-\frac{3}{8}\tilde{q}^{2}+\frac{1}{2}\tilde{r}_{x}+\frac{1}{2}
\langle\tilde{\Phi}_{1},\tilde{\Phi}_{1}\rangle\\
 \end{array}\right)\\
&\tilde{\varphi}_{1jx}=-\tilde{r}\tilde{\varphi}_{1j}+\lambda_{j}\tilde{\varphi}_{2j},
~\tilde{\varphi}_{2jx}=-\lambda_{j}\tilde{\varphi}_{1j}
+\tilde{q}\tilde{\varphi}_{1j}+\tilde{r}\tilde{\varphi}_{2j}+\frac{\mu_{j}}
{\lambda_{j}\tilde{\varphi}_{1j}^{3}}.
\end{align}
\end{subequations}
Since the Kernel of $\tilde{B}_{1}$ is
$(\tilde{c}_{1},\tilde{c}_{2})^{T},$ let
$$-\frac{1}{2}\tilde{q}_{x}-\tilde{q}\tilde{r}+\langle\tilde{\Phi}_{1},\tilde{\Phi}_{2}\rangle=\tilde{c}_{1},
~-\frac{1}{2}\tilde{r}^{2}-\frac{3}{8}\tilde{q}^{2}+\frac{1}{2}\tilde{r}_{x}+\frac{1}{2}
\langle\tilde{\Phi}_{1},\tilde{\Phi}_{1}\rangle=\tilde{c}_{2},
$$
$$\tilde{q}_{1}=\tilde{q},~\tilde{p}_{1}=-\frac{1}{2}\tilde{r},~\tilde{R}=(\tilde{\Phi}_{1}^{T},
\tilde{q}_{1},\tilde{\Phi}_{2}^{T},\tilde{p}_{1},\tilde{c}_{1},\tilde{c}_{2})^{T},$$
then RD-MJMSCS (\ref{mj1}) can be written as a t-type Hamiltonian
system
\begin{subequations}
\label{mj2}
\begin{align}
&\tilde{R}_{x}=\tilde{G}_{1}\frac{\delta \tilde{F}_{1}}{\delta \tilde{R}}\\ \nonumber  where~~~~~~~~~&\\
\nonumber
&\tilde{F}_{1}=-2\tilde{p}_{1}\tilde{c}_{1}+\tilde{q}_{1}\tilde{c}_{2}+2\tilde{p}_{1}^{2}\tilde{q}_{1}
+\frac{1}{8}\tilde{q}_{1}^{3}+2\tilde{p}_{1}\langle
\tilde{\Phi}_{1},\tilde{\Phi}_{2}\rangle-\frac{1}{2}\tilde{q}_{1}\langle
\tilde{\Phi}_{1},\tilde{\Phi}_{1}\rangle\\
&~~~~~~+\frac{1}{2}\langle
\Lambda\tilde{\Phi}_{2},\tilde{\Phi}_{2}\rangle+\frac{1}{2}\langle
\Lambda\tilde{\Phi}_{1},\tilde{\Phi}_{1}\rangle+\sum_{j=1}^{N}\frac{\mu_j}{2\lambda_{j}\tilde{\varphi}_{1j}^{2}},
\\&
G_{1}=\left(\begin{array}{cccc} 0 & I_{(N+1)\times(N+1)}
&0 &0\\
-I_{(N+1)\times(N+1)}& 0 &0& 0\\ 0&0&2\partial_{t}&0\\
0&0&0&\frac{1}{2}\partial_{t}
 \end{array}\right).
\end{align}
\end{subequations}

The Miura map relating systems (\ref{j7}) and (\ref{mj2}), i.e.
$R=M(\tilde{R})$, is given by
\begin{subequations}
\label{hj1}
\begin{align}
&
\Phi_{1}=\tilde{\Phi}_{1},~\Phi_{2}=\Lambda\tilde{\Phi}_{2}+2\tilde{p}_{1}\tilde{\Phi}_{1},\\
&q_{1}=\tilde{q}_{1},~p_{1}=-\frac{1}{2}\tilde{q}_{1}\tilde{p}_{1}-\frac{1}{4}\langle
\tilde{\Phi}_{1},\tilde{\Phi}_{2}\rangle+\frac{1}{4}\tilde{c}_{1},\\
&c_{1}=\frac{1}{2}\tilde{F}_{1}+\partial_{t}\tilde{p}_{1},~~c_{2}=\tilde{c}_{2}.
\end{align}
\end{subequations}
Denote $$M'\equiv \frac{DR}{D\tilde{R}^{T}}$$ where
$\frac{DR}{D\tilde{R}^{T}}$ is the Jacobi matrix consisting of
Frechet derivative of $M$, $M^{'*}$ denotes adjoint of  $M'$.
According to the standard procedure\cite{h6}, applying the map $M$
(\ref{hj1}) to the first Hamiltonian structure of Eq.(\ref{mj2}), we
can generates the second Hamiltonian structure of Eq.(\ref{j7})
\begin{equation}
\label{hj2} G_{2}=M\tilde{G}_{1}M^{*}=\left(\begin{array}{cccccc} 0
& 0
&\Lambda &-\frac{1}{4}\Phi_{1}&\frac{1}{2}\Phi_{2}&0\\
0& 0 &2\Phi_{1}^{T}& -\frac{1}{2}q_{1}&-4p_{1}-\partial_{t}&0\\ -\Lambda&2\Phi_{1}
&0&\frac{1}{4}\Phi_{2}&g_{35}&0\\
\frac{1}{4}\Phi_{1}^{T}&\frac{1}{2}q_{1}&-\frac{1}{4}\Phi_{2}^{T}&\frac{1}{8}\partial_{t}&
g_{45}&0\\
-\frac{1}{2}\Phi_{2}^{T}&4p_{1}-\partial_{t}&-g_{35}&-g_{45}&g_{55}&\partial_{t}q_{1}\\
0&0&0&0&q_{1}\partial_{t}&2\partial_{t}
 \end{array}\right)
\end{equation}
where$$g_{35}=\frac{1}{2}q_{1}\Lambda
\Phi_{1}-\frac{1}{2}\Lambda^{2} \Phi_{1}-\frac{3}{8}
q^{2}_{1}\Phi_{1}-c_{2}\Phi_{1}+\frac{1}{4}\Phi_{1}\langle\Phi_{1},\Phi_{1}\rangle+
(\frac{\mu_{1}}{\varphi_{1j}^{3}},\cdots,\frac{\mu_{N}}{\varphi_{1N}^{3}})^{T}$$
$$g_{45}=-\frac{1}{2}c_{1}+\frac{1}{4}\langle\Lambda\Phi_{1},\Phi_{1}\rangle-\frac{3}{8}q_{1}\langle
\Phi_{1},\Phi_{1}\rangle+\frac{1}{2}q_{1}c_{2}+\frac{1}{8}q_{1}^{3}$$
$$g_{55}=\partial_{t}(\frac{1}{4}
\langle\Phi_{1},\Phi_{1}\rangle-\frac{1}{2}c_{2})+(\frac{1}{4}
\langle\Phi_{1},\Phi_{1}\rangle-\frac{1}{2}c_{2})\partial_{t}-\frac{1}{4}q_{1}\partial_{t}q_{1}.$$
Thus we get the bi-Hamiltonian structure for
Eq.(\ref{j7a})-(\ref{j7c})
\begin{equation}
\label{hj3} R_{x}=G_{1}\frac{\delta F_{1}}{\delta
R}=G_{2}\frac{\delta F_{0}}{\delta R},~~F_{0}=2c_{1}.
\end{equation}

\section{Conclusion}
It is known that there were some methods to construct a new
integrable system starting from a bi-Hamiltonian system. The main
purpose of this paper is to propose the generalized Kupershmidt
deformation (GKD) of bi-Hamiltonian systems to construct new systems
from integrable bi-Hamiltonian systems which is conjectured to be
integrable. We have not be able to prove the integrability of the
generalized Kupershmidt deformation of bi-Hamiltonian systems in
general. However, for many specific cases, such as for  KdV
equation, Boussinesq equation, Jaulent-Miodek equation and
Camassa-Holm equation, by using this generalized Kupershmidt
deformation  some new integrable systems are derived from integrable
bi-Hamiltonian systems. We present a general procedure to convert
this generalized Kupershmidt deformation of the bi-Hamiltonian
systems into an integrable Rosochatius deformation of soliton
equation with self-consistent sources, as well as to transform it
into a $t$-type bi-Hamiltonian system. These imply that the
generalized Kupershmidt deformation of bi-Hamiltonian systems
provides a way to construct new integrable system from an integrable
bi-Hamiltonian systems. On other hand the generalized Kupershmidt
deformation of bi-Hamiltonian systems  also offers a new method to
obtain the integrable Rosochatius deformation of soliton equation
with self-consistent sources, which is quite different from the
method used before. In the further we will continue to study the
integrability of the generalized Kupershmidt deformation of
bi-Hamiltonian systems in general. We believe that the method in [6]
is helpful for proving the bi-Hamiltonian structure of the
generalized Kupershmidt deformation.

\section*{Acknowledgement}
This work is supported by National Basic Research Program of China
(973 Program) (2007CB814800), National Natural Science Foundation of
China (10801083,10901090) and Chinese Universities Scientific Fund
(2009JS42,2009-2-05).


\begin{thebibliography}{10}
\bibitem{ff} B. Fuchssteiner and A. S. Fokas, Physica {\bf4D}  47
(1981).
\bibitem{OR} Peter J. Olver and Philip Rosenau, Phys. Rev. E {\bf53}
1900 (1996).
\bibitem{kdv6}A. Karasu-Kalkani, A. Karasu, A. Sakovich, S. Sakovich and R. Turhan,
J. Math. Phys. {\bf49} 073516 (2008).
\bibitem{kd} Boris A. Kupershmidt, Phys. Lett. A  {\bf372} 2634 (2008).
\bibitem{yz0}  Y. Q. Yao  and  Y. B. Zeng, Lett. Math. Phys.
{\bf86} 193 (2008).
 \bibitem{kkv}  P. H. M. Kersten, I. S. Krasil'shchik, A. M. Verbovetsky and
 R. Vitolo, Acta. Appl. Math. {\bf 109} 75 (2010).
 \bibitem{a1}E.  Rosochatius,   dissertation, University of G$\ddot{o}$tingen,
1877.
\bibitem{a3} S. Wojciechowski ,  Phys. Scr.{\bf 34}  304 (1986).
\bibitem{a4}  R. Kubo, W. Ogura,  T. Saito  and  Y. Yasui,   Phys. Lett. A  {\bf 251} 6 (1999).
\bibitem{a5} R. G. Zhou  , J. Math. Phys.
  {\bf48} 103510 (2007).
\bibitem{a6} C. Bartocci,  G. Falqui  and M. Pedroni,  Diff. Geom. Applic. {\bf 21}
349 (2004).
\bibitem{a7}  M. Kruczenski,  J. G. Russo  and  A. A. Tseytlin,  J. High Energy Phys.
{\bf10}  063 (2006).
\bibitem{a8} P. L. Christiansen, J. C. Eilbeck, V. Z. Enolskii  and N. A. Kostov,  Proceedings:
Mathematical and Physical Sciences.{\bf 451}  685 (1995).
\bibitem{yz}    Y. Q. Yao  and  Y. B. Zeng, J. Phys. A: Math. Theor.
 {\bf41} 295205 (2008) .
 \bibitem{yz1} Y. Q. Yao and Y. B. Zeng, Commun. Theor. Phys. {\bf52} 193 (2009).
 \bibitem{yz2} V. I. Arnold, Mathematical Methods of Classical
Mechanics,  Spriger-Verlag, New York (1978).
\bibitem{h1} B. Fuchssteiner , W. Oevel,
Physica A  {\bf 145}  67 (1987).
\bibitem{h2} A. P. Fordy,
 Physica D {\bf 87}  20 (1995).
\bibitem{h3} M. Blaszak,
 J. Math. Phys. {\bf 36}  4826 (1995).
\bibitem{h4}Y. B. Zeng,
 Physica A {\bf 262} 405 (1999).
\bibitem{ks} V. K.  Menlikov, Inverse Probl. {\bf 6} 233 (1990).
\bibitem{ch1} R. Camassa and D. Holm,  Phys. Rev. Lett.  {\bf 71}  1661 (1993).
\bibitem{ch3}
 A. Parker,
 Proc. R. Soc. Lond. A  {\bf 460}   2929 (2004).
 \bibitem{yz4}
 Y. H. Huang, Y. Q. Yao and Y. B. Zeng. , Commun. Theor. Phys. {\bf 53} 403
 (2010).
\bibitem{b1}
A. P. Fordy and J. Gibbons, J. Math. Phys.   {\bf 22} 1170 (1981).
\bibitem{b2}
Y. B. Zeng, Acta Mathematica Scientia {\bf 17} 97 (1997).
\bibitem{j1}
 M. Jaulent and K. Miodek.
  Lett. Math. Phys. {\bf 1}  243  (1976).
\bibitem{Fordy}
A. P. Fordy, Isospectral flows: their Hamiltonian structure, Miura
maps and master symmetries, in: P. J. Olver, D. H.
Sattinger(Eds.),Proc. IWA Workshop on Applications of Solitons,
1991, preprint.
\bibitem{h6}
B. A. Kupershmidt and  J. Wilson, Invent Math.{\bf 62} 403 (1981).



\end{thebibliography}
\end{document}